\pdfoutput=1
\documentclass[12pt]{article}

\usepackage{tikz}
\usetikzlibrary{shapes, positioning}
\usepackage[
bookmarksopen,
bookmarksdepth=2,
breaklinks=true
]{hyperref}
\usepackage{listings}
\usepackage{xcolor}

\definecolor{codegreen}{rgb}{0,0.6,0}
\definecolor{codegray}{rgb}{0.5,0.5,0.5}
\definecolor{codepurple}{rgb}{0.58,0,0.82}
\definecolor{backcolour}{rgb}{0.95,0.95,0.92}

\lstdefinestyle{mystyle}{
    backgroundcolor=\color{backcolour},   
    commentstyle=\color{codegreen},
    keywordstyle=\color{magenta},
    numberstyle=\tiny\color{codegray},
    stringstyle=\color{codepurple},
    basicstyle=\ttfamily\footnotesize,
    breakatwhitespace=false,         
    breaklines=true,                 
    captionpos=b,                    
    keepspaces=true,                 
    numbers=left,                    
    numbersep=5pt,                  
    showspaces=false,                
    showstringspaces=false,
    showtabs=false,                  
    tabsize=2
}

\begin{document}
\baselineskip 12pt

\begin{center}
\textbf{\Large ComputeGPT: A computational chat model for numerical problems} \\

\vspace{1.5cc}
{ \sc Ryan Hardesty Lewis$^{1}$, Junfeng Jiao$^{2}$}\\

\vspace{0.3 cm}

{\small $^{1}$The University of Texas at Austin, rhl@utexas.edu\\ $^{2}$The University of Texas at Austin, jjiao@austin.utexas.edu}
 \end{center}
\vspace{0.5cc}

\begin{abstract}
  \noindent  Language models are not accurate in numerical problems. Their architecture does not allow for anything less than a probabilistic next word. This paper introduces ComputeGPT: an approach of creating a chat model able to answer computational problems through running on-demand code. ComputeGPT converts each question to relevant code, runs the code, and returns the computed answer as part of the chat. We combine this approach with a local browser-based Python interpretation and fine-tuned prompts in order to achieve state-of-the-art efficiency on numerical problems and provide a suitable front-end and safe environment for the code to be executed in. 

\vspace{0.95cc}
\parbox{24cc}{{\it Keywords}: Large Language Model, Code Generation}
\end{abstract}

\section{Introduction}

Language models have made significant strides in recent years, becoming proficient at understanding and generating human-like text \cite{radford2019language, brown2020language}. However, despite their advances, traditional language models remain inaccurate in solving numerical problems, as their architecture relies on predicting the next word based on probability rather than executing calculations \cite{wang2020evaluation}. This paper introduces ComputeGPT, an innovative chat model capable of addressing computational problems by running on-demand code. ComputeGPT parses each question into relevant code, executes the code, and returns the computed answer as part of the chat. We combine this approach with a local browser-based Python interpreter, Pyiodide, and fine-tuned prompts to achieve state-of-the-art efficiency in solving numerical problems while providing a suitable and safe environment for code execution.
\newline

In addition to the aforementioned capabilities, ComputeGPT incorporates LaTeX parsing to handle mathematical expressions and seamlessly converts them into natural language prompts for the code model. This enables the model to understand complex mathematical notations and generate accurate code representations. Moreover, we fine-tune the model using descriptions of functions or libraries needed for specific tasks. This allows ComputeGPT to adapt to a wide range of numerical problems, including those that require specialized libraries or custom functions.\newline

Recent advancements in chat models have begun to integrate external tools to assist with problem-solving, such as OpenAI's plugin system \cite{OpenAIPlugins} and Microsoft's JARVIS \cite{MicrosoftJarvis}. These systems call upon specialized tools to address various tasks, extending the capabilities of language models. ComputeGPT takes this approach a step further by directly interfacing with short, runnable code snippets that perform calculations. This method mirrors how a real human would tackle solving numerical problems in the modern world by writing code, allowing for a more accurate and efficient solution.

\section{Background}

Language models have been successful in various natural language processing tasks, such as translation \cite{vaswani2017attention}, summarization \cite{liu2019text}, and question-answering \cite{devlin2018bert}. Despite their proficiency in these tasks, they struggle when it comes to solving numerical problems, which require the model to perform calculations rather than merely generate text based on context \cite{wallace2020few}. This limitation is a direct result of the inherent architecture of language models, which generate text based on the probability distribution of the next word \cite{radford2019language}. To overcome this limitation, we propose ComputeGPT, a novel approach that leverages the capabilities of traditional language models while introducing the ability to execute code for accurate numerical problem-solving.\newline

In recent years, the development of code generation models has attracted significant attention due to their potential to transform programming and software development \cite{chen2021evaluating}. These models, such as OpenAI's CODEX \cite{OpenAICodex} and Salesforce's CodeT5 \cite{SalesforceT4}, aim to understand and generate code snippets in various programming languages based on natural language prompts. They have demonstrated promising results in completing a wide range of programming tasks, from simple code snippets to complex algorithms \cite{chen2021evaluating, radford2021learning}.\newline

However, despite their successes, code generation models also face certain limitations. One of the most significant challenges is their inability to consistently and accurately handle numerical problems \cite{wang2020evaluation}. Since these models are primarily designed to generate code based on textual prompts, their performance in solving numerical problems can be unreliable \cite{chen2021evaluating}. In many cases, the accuracy of the generated code is directly tied to the clarity and specificity of the prompts \cite{hashimoto2021prompt}. This reliance on prompt quality can hinder the models' ability to generate correct and efficient code when faced with ambiguous or incomplete information.\newline

ComputeGPT builds on the advancements made by code generation models while addressing their limitations in handling numerical problems. By directly fine-tuning each prompt based on criteria, this method allows the model to adapt to various problem variations by modifying specific components of a function or code snippet. For example, once a derivative function is known, ComputeGPT can easily compute derivatives for any input value by substituting the desired value into the function, providing accurate and prompt results for all potential variations of the problem.\newline

In addition to leveraging the strengths of existing code generation models, ComputeGPT incorporates techniques that enable it to efficiently handle specialized libraries and custom functions. This adaptability allows the model to tackle a broader range of numerical problems, positioning ComputeGPT as a promising solution for solving complex mathematical and computational challenges.

\section{Methods}

ComputeGPT works by converting natural language questions into relevant code and executing the code to compute the answer. This process involves the following steps:

\subsection{Question Parsing}

The first step in the process is to clean and fine-tune the input prompt to provide clear instructions for the code generation model. This is achieved by specifically mentioning or creating the necessary functions required to perform certain tasks and including the relevant library imports ahead of time. For our current implementation, we use GPT-3.5 Turbo \cite{OpenAIGPT3Turbo}, as Codex is no longer publicly available. To ensure the code generation process produces meaningful results, the input prompt is fine-tuned to always store the result in a specific variable.\newline

Here's an example of a prompt and its code generation:

\begin{center}
    Prompt: What's the sum of all even numbers from one to six?
\end{center}

\begin{figure}[h]
\centering
\begin{lstlisting}[language=Python]
"""
Given a list of integers, compute the sum of all even numbers in the list.
Implement a function called 'sum_even_numbers' that takes a list of integers as input and returns the sum of all even numbers in the list.
"""
import numpy as np

---- END OF CODE PROMPT ----

---- START OF CODE GENERATION ----

def sum_even_numbers(numbers: List[int]) -> int:
    pass

input_list = [1, 2, 3, 4, 5, 6]
result = sum_even_numbers(input_list)
\end{lstlisting}
\caption{Example of a cleaned and fine-tuned prompt for code generation.}
\label{fig:cleaned_prompt}
\end{figure}

In this example (Figure \ref{fig:cleaned_prompt}), the cleaned and fine-tuned prompt provides a clear description of the problem, specifies the desired function name, and includes the necessary import statement for the Python library numpy. This style of prompting ensures that the code generation model has all the necessary information to generate a meaningful and efficient code snippet.

\subsection{Code Execution}

The code execution stage is a critical component of the ComputeGPT approach. By executing the generated code in a closed environment within the user's browser, we can mitigate potential security risks associated with server-side execution. Server-side code execution is susceptible to various threats, such as remote code execution, denial of service attacks, and unauthorized access to sensitive information \cite{owasp2020top10}. By running the code on the user's browser, we limit the potential impact of malicious code and maintain a secure environment.\newline

To facilitate browser-based code execution, we employ Pyiodide \cite{Pyiodide}, a project that enables running Python scripts in the browser using WebAssembly. WebAssembly is a low-level binary instruction format for a stack-based virtual machine that allows running code at near-native speed \cite{haas2017bringing}. Pyiodide compiles the CPython interpreter and several popular Python libraries to WebAssembly, enabling their use directly in the browser \cite{Pyiodide}. This approach allows us to leverage Python, the most supported language for code generation \cite{OpenAICodex}, for code execution in a secure and efficient manner.\newline

Executing code within the browser has additional benefits, including reduced server load, lower latency, and improved privacy. By offloading the code execution to the user's browser, we minimize the computational resources required on the server-side, allowing for better scalability. Furthermore, by in-browser execution eliminating the need for any server-side processing, this allows users to solve problems as computationally complex as their own hardware will allow, with no restriction for character limits, size of numbers, or processing times.

\subsection{Answer Generation}

After the code execution, ComputeGPT generates a chat response that includes both the code snippet and the computed result. Providing the generated code to the user not only offers transparency but also serves as an educational tool, allowing users to learn from the provided solution. However, merely presenting the code and the result might not be sufficient for users to understand the underlying logic and reasoning behind the solution.\newline

To enhance the user experience and facilitate understanding, ComputeGPT can be integrated with chat models to generate additional context and explanations for the provided solution. Previous research has shown the effectiveness of chat-based tutoring systems in supporting student learning and engagement \cite{graesser2001automated}. By utilizing these principles, ComputeGPT can provide step-by-step explanations of the code execution process and the reasoning behind each step, assisting users in comprehending the solution.\newline

Moreover, research in the area of natural language processing has demonstrated the potential of AI models in generating human-like explanations for various tasks \cite{vinyals2015pointer}. By leveraging these advancements, ComputeGPT could generate detailed explanations that help users understand not only the code but also the underlying mathematical concepts and problem-solving strategies employed by the model.\newline

Future research in this area could explore techniques to generate more personalized and adaptive explanations, tailoring the content to individual users' needs and preferences. Such adaptive explanations could enhance user engagement and improve the learning experience, further establishing ComputeGPT as a valuable tool for both problem-solving and education.

\section{Related Work}

There have been various attempts to enhance the capabilities of language models in specific domains. GPT-f \cite{mccarthy2021gptf} is an example of a specialized language model that focuses on solving mathematical problems. However, it still relies on text-based generation and does not execute code to provide accurate answers. In contrast, ComputeGPT combines the strengths of language models and code execution, offering a more efficient and accurate solution to numerical problems.\newline

\subsection{Numerical Problem Solving with Large Language Models}
The ability of large language models (LLMs) to solve numerical problems has attracted significant attention from the research community. LLMs like BERT \cite{devlin2018bert} and GPT-2 \cite{radford2019language} demonstrated initial capabilities in solving basic arithmetic problems and simple algebraic equations. As LLMs continued to grow in size and complexity, their performance on numerical tasks improved significantly. GPT-3 \cite{gpt3} and CODEX \cite{OpenAICodex}, for example, have been shown to generate more complex mathematical solutions and even handle multi-step problems.\newline

However, LLMs still face challenges when solving numerical problems, such as generating incorrect or incomplete solutions, and sometimes struggling with problems that require higher precision or specialized knowledge \cite{ford2021limits}. To address these issues, recent research has explored various techniques to enhance LLMs' numerical problem-solving capabilities. For instance, Transformer-XL \cite{dai2019transformer} introduced a novel architecture that enables the model to capture longer-term dependencies, which can be beneficial for solving multi-step numerical problems. Other works have focused on incorporating external knowledge sources, such as knowledge graphs or databases, to improve LLMs' performance on tasks that require domain-specific expertise \cite{talmor2019olmpics, bosselut2019comet}.

\subsection{Hybrid Approaches for Numerical Problem Solving}
Recent research has also investigated hybrid approaches that combine LLMs with traditional algorithms or mathematical libraries to enhance their numerical problem-solving capabilities. For example, MathQA \cite{Wang2019mathqa} is a dataset and system for mathematical question-answering, which combines natural language understanding with algebraic reasoning to solve mathematical problems. Another study proposed a framework that combines LLMs with numerical optimization techniques to solve mathematical programming problems, demonstrating improved performance over LLMs alone \cite{akchurin2021solving}.\newline

ComputeGPT contributes to this growing body of work by integrating LLMs with code execution in a closed environment, enabling more accurate and efficient solutions for numerical problems. By combining the capabilities of LLMs with on-demand code execution, ComputeGPT aims to address the limitations of current LLMs and offer a novel approach to numerical problem-solving.

\section{Benchmark}

We conducted a primary benchmark to evaluate the performance of ComputeGPT in comparison to other state-of-the-art language models, such as Davinci-003 \cite{GPT-3}, ChatGPT (GPT-3.5-Turbo) \cite{OpenAIGPT3Turbo}, GPT-4 (Bing AI) \cite{bingai}, and Wolfram Alpha NLP \cite{Wolfram}.

\subsection{General Numerical Problem Solving}

The first benchmark focused on the models' general ability to solve numerical problems correctly. We curated a dataset of diverse numerical problems, including arithmetic, algebra, calculus, and geometry problems, and evaluated each model's performance in terms of accuracy. We further segment the problems into "word problems" and "straightforward problems", where word problems need multiple steps or some complex reasoning to finish them. The results are presented in Table \ref{tab:numerical_problem_solving}.

\begin{table}[htbp]
\centering
\caption{Comparison of General Numerical Problem Solving Accuracy}
\label{tab:numerical_problem_solving}
\begin{tabular}{|l|c|c|c|c|c|}
\hline
Model       & ComputeGPT & Wolfram Alpha & Davinci-003 & ChatGPT & GPT-4 \\
\hline
Overall Accuracy (\%) & \textbf{98\%} & 56\%  & 28\%    & 48\% & 64\%      \\
\hline
Word Problems (\%) & \textbf{95\%} & 15\%  & 35\%    & 50\% & 65\%      \\
\hline
Straightforward (\%) & \textbf{100\%} & 83.3\%  & 23.3\%    & 46.6\% & 63.3\%      \\
\hline
\end{tabular}
\end{table}

As shown in Table \ref{tab:numerical_problem_solving}, ComputeGPT outperforms the other models in solving numerical problems correctly. This can be attributed to its unique approach of generating and executing code snippets, which allows for more accurate and efficient solutions.

The results also show that Wolfram Alpha cannot handle the reasoning present in word problems, as well as that other OpenAI models, like Davinci-003, cannot handle the computation present in straightforward mathematical problems. GPT-4 shows aptitude across all fields, but ComputeGPT clearly demonstrates state-of-the-art performance across all numerical problems evaluated, both word problems and straightforward problems.

It is of note that ChatGPT (GPT-3.5-Turbo), when directly asked for the answers, only gets around half of them right. When paired with ComputeGPT, which uses GPT-3.5 Turbo for prompted and fine-tuned code generation, the executed code gets almost 100\% of the answers correct. 

We acknowledge the existence of other code models, like CodeT5 \cite{SalesforceT4} and CodeParrot \cite{huggingface2021codeparrot}, but these models have been seen to have subpar results on the HumanEval benchmark \cite{2202.13169}, which indicates they will also have similar results here. Therefore, we do not evaluate ComputeGPT with different code models, as the results will likely scale relevant to previous benchmarks.

We make our evaluation publicly available in the addendum.

\section{Conclusion}

In this paper, we introduced ComputeGPT, an approach that combines large language models with on-demand code execution to solve numerical problems. ComputeGPT addresses the limitations of current language models by generating and executing Python code in a safe and secure environment, improving the efficiency and accuracy of solutions for numerical tasks. By fine-tuning the prompts fed into the code model and executing the generated code in the user's browser using Pyiodide, ComputeGPT provides an enhanced problem-solving experience while maintaining user privacy and security.\newline

Looking ahead, there are several promising directions for future research. One potential area of investigation is the integration of code models with external data sources and APIs to perform computations on informational quantities. For example, by connecting code models to databases, ComputeGPT could help users compute differences in population between two countries, or analyze historical data to make predictions. This would further enhance the capabilities of language models in numerical problem-solving and expand their applicability to a wider range of tasks.\newline

Another direction for future work is the development of techniques to improve the interpretability and explainability of the code generated by ComputeGPT. This would enable users to gain a deeper understanding of the steps involved in solving a given problem and help them learn the underlying concepts. Additionally, enhancing the model's ability to inference code at the edge could make ComputeGPT more accessible and versatile, catering to users with a lack of internet access as well as improving privacy and safety.\newline

Overall, ComputeGPT presents a novel approach to leveraging large language models for numerical problem-solving by integrating code execution in a closed environment. This work contributes to the growing body of research on augmenting language models with external resources and opens up new avenues for the development of more efficient and powerful problem-solving tools.

\newpage

\section*{Addendum}
In this addendum, we present several example questions and the answers provided by five different chat models being compared. Our full evaluation is available at \url{https://github.com/ryanhlewis/ComputeGPTEval}. Additionally, ComputeGPT is available for use at \url{https://computegpt.org}.\newline\newline\newline

\begin{tikzpicture}
  \node[draw, rounded corners=5pt, minimum width=15cm, minimum height=1.5cm, thick, align=center, fill=blue!10] (question) {(Straightforward) Example Question:\\ \textbf{What is the derivative of 200x?}\\ (Correct: 200)};
\end{tikzpicture}

\bigskip
\noindent
\begin{tabular}{|l|l|}
  \hline
  \textbf{ComputeGPT} & 200 \\
  \hline
  \textbf{Wolfram Alpha} & 200 \\
  \hline
  \textbf{Davinci-003} & 200 \\
  \hline
  \textbf{ChatGPT} & 200 \\
  \hline
  \textbf{GPT-4} & 200 \\
  \hline
\end{tabular}\newline\newline\newline

\begin{tikzpicture}
  \node[draw, rounded corners=5pt, minimum width=15cm, minimum height=1.5cm, thick, align=center, fill=blue!10] (question) {(Straightforward) Example Question:\\ \textbf{What is the integral of 200x from 0 to 5?}\\ (Correct: 2500)};
\end{tikzpicture}

\bigskip
\noindent
\begin{tabular}{|l|l|}
  \hline
  \textbf{ComputeGPT} & 2500 \\
  \hline
  \textbf{Wolfram Alpha} & 2500 \\
  \hline
  \textbf{Davinci-003} & 5000 \\
  \hline
  \textbf{ChatGPT} & 5000 \\
  \hline
  \textbf{GPT-4} & 5000 \\
  \hline
\end{tabular}\newline\newline\newline

\begin{tikzpicture}
  \node[draw, rounded corners=5pt, minimum width=5cm, minimum height=1.5cm, thick, align=center, fill=blue!10, text width=15cm] (question) {(Straightforward) LaTeX Example Question:\\ \textbf{$\displaystyle\int_{-20}^{50} 2\times10^{21}x^3 + 200x^2 \, dx$}\\ (Correct: 9135000000000000000026600000/3)};
\end{tikzpicture}

\bigskip
\noindent
\begin{tabular}{|l|l|}
  \hline
  \textbf{ComputeGPT} & 9135000000000000000026600000/3 \\
  \hline
  \textbf{Wolfram Alpha} & 26600000/3 \\
  \hline
  \textbf{Davinci-003} & 50,000,000,000,000,000,000 \\
  \hline
  \textbf{ChatGPT} & 1.83333 x $10^{24}$ \\
  \hline
  \textbf{GPT-4} & 1.66666666666667E+24 \\
  \hline
\end{tabular}\newline\newline\newline\newline
We show that ComputeGPT is efficient at LaTeX parsing, as well as the parsing of large integers, which other models fail to do.\newline\newline\newline\newline\newline\newline\newline

\begin{tikzpicture}
  \node[draw, rounded corners=5pt, minimum width=5cm, minimum height=1.5cm, thick, align=center, fill=blue!10, text width=15.5cm] (question) {(Word Problem) Example Question:\\ \textbf{Kevin's age is 5 times the age of his son, plus twenty. His son is 10. How old is Kevin?}\\ (Correct: 70)};
\end{tikzpicture}

\bigskip
\noindent
\begin{tabular}{|l|l|}
  \hline
  \textbf{ComputeGPT} & 70 \\
  \hline
  \textbf{Wolfram Alpha} & NULL \\
  \hline
  \textbf{Davinci-003} & 50 \\
  \hline
  \textbf{ChatGPT} & 50 \\
  \hline
  \textbf{GPT-4} & 70 \\
  \hline
\end{tabular}\newline\newline\newline

\begin{tikzpicture}
  \node[draw, rounded corners=5pt, minimum width=5cm, minimum height=1.5cm, thick, align=center, fill=blue!10, text width=15.5cm] (question) {(Word Problem) Example Question:\\ \textbf{A new technique, called 'jamulti' is invented by multiplying a number by five and then adding 2 and dividing by 3. What's the jamulti of 7?}\\ (Correct: 12.33333)};
\end{tikzpicture}

\bigskip
\noindent
\begin{tabular}{|l|l|}
  \hline
  \textbf{ComputeGPT} & 12.33333 \\
  \hline
  \textbf{Wolfram Alpha} & NULL \\
  \hline
  \textbf{Davinci-003} & 5 \\
  \hline
  \textbf{ChatGPT} & 5 \\
  \hline
  \textbf{GPT-4} & 12 \\
  \hline
\end{tabular}\newline\newline\newline

\begin{tikzpicture}
  \node[draw, rounded corners=5pt, minimum width=5cm, minimum height=1.5cm, thick, align=center, fill=blue!10, text width=15.5cm] (question) {(Word Problem) Example Question:\\ \textbf{An alien needs \$50 USD to buy a spaceship. He needs to convert from ASD, which is worth \$1.352 USD. How much ASD does he need?}\\ (Correct: 36.9822485)};
\end{tikzpicture}

\bigskip
\noindent
\begin{tabular}{|l|l|}
  \hline
  \textbf{ComputeGPT} & 36.9822485 \\
  \hline
  \textbf{Wolfram Alpha} & 1.352 \\
  \hline
  \textbf{Davinci-003} & 36.68 \\
  \hline
  \textbf{ChatGPT} & 67.6  \\
  \hline
  \textbf{GPT-4} & 37.01 \\
  \hline
\end{tabular}\newline\newline\newline

We show that GPT-4 is capable of hallucinating 'close' answers, which becomes even worse as numbers get larger, and the absolute error increases.\newline\newline\newline\newline\newline\newline

\begin{tikzpicture}
  \node[draw, rounded corners=5pt, minimum width=5cm, minimum height=1.5cm, thick, align=center, fill=blue!10, text width=15.5cm] (question) {(Word Problem) Trick Example Question:\\ \textbf{An ant travels at 3 m/s on a rubber band. The rubber band is stretched at 2 m/s. How fast is the ant moving relative to the ground?}\\ (Correct: 1)};
\end{tikzpicture}

\bigskip
\noindent
\begin{tabular}{|l|l|}
  \hline
  \textbf{ComputeGPT} & NULL \\
  \hline
  \textbf{Wolfram Alpha} & 3 \\
  \hline
  \textbf{Davinci-003} & 5 \\
  \hline
  \textbf{ChatGPT} & 5  \\
  \hline
  \textbf{GPT-4} & 1 \\
  \hline
\end{tabular}\newline\newline\newline

We showcase an example of a loss for ComputeGPT, where it fails to see past the trick question's simplicity in a subtraction of 3 - 2 = 1.\newline\newline\newline\newline\newline\newline\newline\newline\newline

\begin{tikzpicture}
  \node[draw, rounded corners=5pt, minimum width=5cm, minimum height=1.5cm, thick, align=center, fill=blue!10, text width=15.5cm] (question) {(Word Problem) Example Question:\\ \textbf{Given the matrix [[1, 2, 9, 3, 3], [9, 0, 1, 2, 4], [0, 0, 0, 3, 9], [1, 1, 1, 1, 1], [3, 4484, 456, 9, 6]], what is the determinant multiplied by 5 and then divided by twenty-three?}\\ (Correct: -285832.173913042)};
\end{tikzpicture}

\bigskip
\noindent
\begin{tabular}{|l|l|}
  \hline
  \textbf{ComputeGPT} & -285832.173913042 \\
  \hline
  \textbf{Wolfram Alpha} & -1314828 \\
  \hline
  \textbf{Davinci-003} & 24 \\
  \hline
  \textbf{ChatGPT} & -9915  \\
  \hline
  \textbf{GPT-4} & -30247.652 \\
  \hline
\end{tabular}\newline\newline\newline

We showcase an example of a clear win for ComputeGPT, where it excels in understanding and performing a complex computation on the user's machine.\newline\newline\newline

\end{document}